\documentclass[sigconf]{acmart}

\usepackage{subcaption}
\usepackage{enumitem}
\usepackage{multirow}
\usepackage{makecell}
\usepackage{threeparttable}
\usepackage{caption}
\AtBeginDocument{%
  }

\copyrightyear{2024}
\acmYear{2024}
\setcopyright{acmlicensed}\acmConference[ICCAD '24]{IEEE/ACM International Conference on Computer-Aided Design}{October 27--31, 2024}{New York, NY, USA}
\acmBooktitle{IEEE/ACM International Conference on Computer-Aided Design (ICCAD '24), October 27--31, 2024, New York, NY, USA}
\acmDOI{10.1145/3676536.3676704}
\acmISBN{979-8-4007-1077-3/24/10}

\begin{document}

%%
%% The "title" command has an optional parameter,
%% allowing the author to define a "short title" to be used in page headers.
\title{PACiM: A Sparsity-Centric Hybrid Compute-in-Memory Architecture via Probabilistic Approximation}

%%
%% The "author" command and its associated commands are used to define
%% the authors and their affiliations.
%% Of note is the shared affiliation of the first two authors, and the
%% "authornote" and "authornotemark" commands
%% used to denote shared contribution to the research.

\author{Wenlun Zhang}
\affiliation{%
 \institution{Keio University}
 \city{Kanagawa}
 \country{Japan}}
\email{wenlun_zhang@keio.jp}

\author{Shimpei Ando}
\affiliation{%
 \institution{Keio University}
 \city{Kanagawa}
 \country{Japan}}
\email{shimpeiando@keio.jp}

\author{Yung-Chin Chen}
\affiliation{%
 \institution{Keio University}
 \city{Kanagawa}
 \country{Japan}}
\email{jim.chen.work@gmail.com}

\author{Satomi Miyagi}
\affiliation{%
 \institution{Keio University}
 \city{Kanagawa}
 \country{Japan}}
\email{stm-m22@keio.jp}

\author{Shinya Takamaeda-Yamazaki}
\affiliation{%
 \institution{The University of Tokyo}
 \city{Tokyo}
 \country{Japan}}
\email{shinya@is.s.u-tokyo.ac.jp}

\author{Kentaro Yoshioka}
\affiliation{%
 \institution{Keio University}
 \city{Kanagawa}
 \country{Japan}}
\email{kyoshioka47@keio.jp}

%%
%% By default, the full list of authors will be used in the page
%% headers. Often, this list is too long, and will overlap
%% other information printed in the page headers. This command allows
%% the author to define a more concise list
%% of authors' names for this purpose.
\renewcommand{\shortauthors}{Zhang et al.}

%%
%% The abstract is a short summary of the work to be presented in the
%% article.
\begin{abstract}
  Approximate computing emerges as a promising approach to enhance the efficiency of compute-in-memory (CiM) systems in deep neural network processing. However, traditional approximate techniques often significantly trade off accuracy for power efficiency, and fail to reduce data transfer between main memory and CiM banks, which dominates power consumption. This paper introduces a novel probabilistic approximate computation (PAC) method that leverages statistical techniques to approximate multiply-and-accumulation (MAC) operations, reducing approximation error by $4\times$ compared to existing approaches. PAC enables efficient sparsity-based computation in CiM systems by simplifying complex MAC vector computations into scalar calculations. Moreover, PAC enables sparsity encoding and eliminates the LSB activations transmission, significantly reducing data reads and writes. This sets PAC apart from traditional approximate computing techniques, minimizing not only computation power but also memory accesses by 50\%, thereby boosting system-level efficiency. We developed PACiM, a sparsity-centric architecture that fully exploits sparsity to reduce bit-serial cycles by 81\% and achieves a peak 8b/8b efficiency of 14.63 TOPS/W in 65 nm CMOS while maintaining high accuracy of 93.85/72.36/66.02\% on CIFAR-10/CIFAR-100/ImageNet benchmarks using a ResNet-18 model, demonstrating the effectiveness of our PAC methodology. \textbf{Software simulation framework is available at} \href{https://github.com/Keio-CSG/PACiM}{\textit{GitHub}}.
\end{abstract}

%%
%% The code below is generated by the tool at http://dl.acm.org/ccs.cfm.
%% Please copy and paste the code instead of the example below.
%%

%%
%% Keywords. The author(s) should pick words that accurately describe
%% the work being presented. Separate the keywords with commas.

%% A "teaser" image appears between the author and affiliation
%% information and the body of the document, and typically spans the
%% page.

%%
%% This command processes the author and affiliation and title
%% information and builds the first part of the formatted document.
\maketitle

\section{Introduction}

\begin{figure*}[htbp]
\centering
\includegraphics[width=\textwidth]{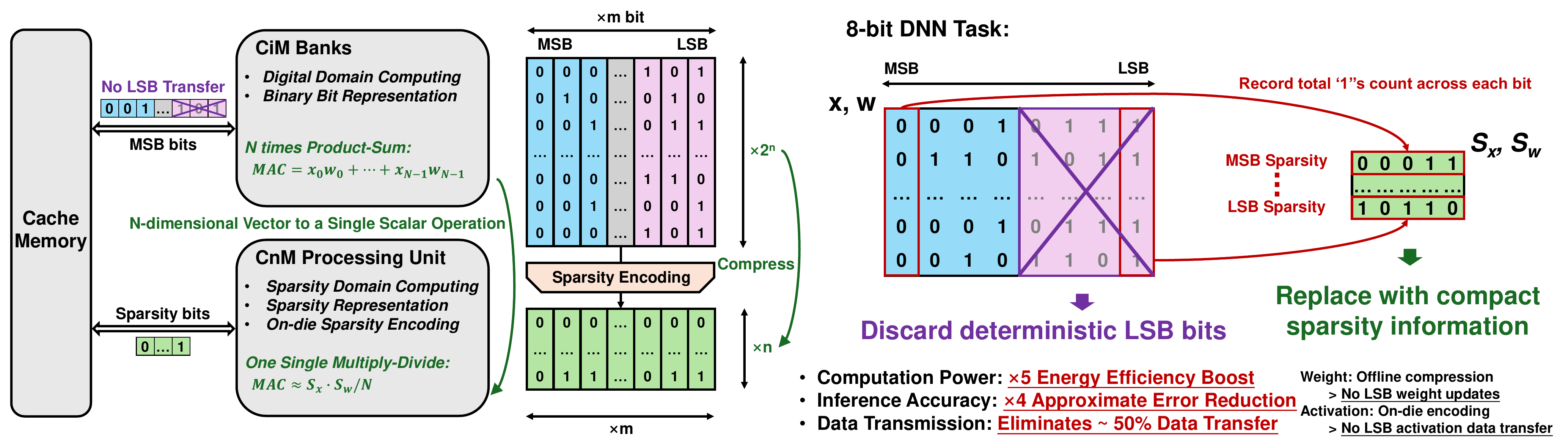}
\caption{General Concept of PACiM: This architecture utilizes a novel PAC method to transform traditional vector computations into a single scalar operation. It features data compression into bit-level sparsity, while eliminating all LSB data transfers.}
\label{Fig_Concept}
\end{figure*}

Deep learning has witnessed remarkable advancements in recent years, extending from image recognition and natural language processing to autonomous systems and predictive analytics\cite{DNN_Processing_Survey}. To address the computational demands of deep neural network (DNN) models, Compute-in-Memory (CiM) architectures have emerged as a promising alternative, marking a paradigm shift away from conventional computing approaches\cite{CiM_Prospects}. By incorporating processing capabilities directly within memory components, CiM architectures minimize data movement through data reuse, overcoming a critical limitation of traditional von Neumann architectures\cite{64-Tile,Bit-Parallel}. However, the increasing complexity of DNN models, often requiring wider operand bit-widths, results in a corresponding increase in computational burden, leading to significant energy costs for DNN applications\cite{Model_Benchmark}. This has driven interest in approximate computing within the CiM domain due to its potential to deliver high energy efficiency while maintaining high precision in DNN inference\cite{Approximate_Survey}. The most prevalent approximate computing architecture is the Hybrid CiM (H-CiM), combining digital and analog computing: it processes deterministic precision computation on the MSB part within the digital domain, while the LSB part is transferred to the energy-efficient analog domain, yielding a substantial energy efficiency gain with minimal accuracy loss\cite{Hybrid_Analog_Digital_CiM}.

Despite the potential of approximate computation to transform CiM architecture, it faces two critical constraints that hinder its broader application. The first constraint involves the trade-off between approximation accuracy and resource overhead. Take hybrid digital-analog computing for example, where the analog domain computing is used to approximate LSB computations. Achieving greater computational accuracy requires higher-precision ADCs, which impacts energy efficiency and adds substantial area overhead. This creates a contradiction where existing approximate methods cannot concurrently achieve high precision and high energy efficiency, presenting a significant challenge for designing high-performance approximate CiM systems.

The second major challenge concerns the substantial energy overhead caused by data movement between memory and CiM macros. Although CiM outperforms conventional digital accelerators at the macro level, this merit is less evident at a system level\cite{CiM_Benchmark}. The main reason for this discrepancy is frequent cache memory accesses for transferring output or intermediate activations, highlighting the need to optimize the activation movement to unlock the full potential of CiM. However, recent approximate methodologies in CiM often focus solely on computation, neglecting the impact of memory access. Extensive research has shown that the energy cost of cache access is comparable to computation, and weight updates requiring off-die DRAM access can be orders of magnitude more expensive\cite{Interstellar,Bit-Parallel,Scalable_Programmable_CiM}. This suggests that approximate methodologies that address memory access are crucial for achieving high system-level energy efficiency.

In this paper, we propose a novel probabilistic approximate computation (PAC) methodology and a sparsity-centric architecture to address the above two challenges. As Fig. \ref{Fig_Concept} demonstrates, the PAC method effectively reduces extensive m-bit, $2^{n}$-length weight and activation data to a compact $m \times n$ bit-level sparsity, transitioning computations into the sparsity domain. Unlike the traditional multiply-and-accumulation (MAC) operations, which execute N times product-sum calculations, computations within the sparsity domain approximate the MAC output through a singular multiply-divide operation, markedly improving computational efficiency. The proposed PACiM architecture combines CiM banks with a Compute-near-Memory (CnM) processing unit into a hybrid system. This design ensures that the MSB portion of the MAC computation remains accurate within the digital domain using conventional binary bit data representation. Meanwhile, the CnM unit employs the PAC methodology for approximate computing using bit-level sparsity, surpassing existing approximate computing techniques in both computational efficiency and accuracy. Through on-die sparsity encoding and data translation into a bit-level sparsity, our method eliminates the LSB activation transmission and facilitates a low-cost sparsity bit communication. This leads to a substantial reduction in expensive DRAM and cache accesses. Our contributions are outlined as follows:

\begin{itemize}[leftmargin=*]
    \item Moving beyond traditional techniques, we find that the PAC method can accurately approximates the MAC output with a root-mean-square error (RMSE) below 1\%, which represents a fourfold improvement over competing approximation methods in the CiM paradigm.
    \item The design of PACiM, a sparsity-centric architecture that leverages the full potential of PAC, reducing bit-serial computational cycles by over 81\% and achieving peak 8b/8b efficiency of 14.63 TOPS/W in a 65 nm CMOS, along with a 50\% reduction in data access power. To our best knowledge, this is the first approximate CiM design that tackles the system-level cache access issue utilizing sparsity encoding.
    \item Extensive experimental validation shows that our PACiM architecture maintains high accuracy levels of 93.85\% (-0.62\%), 72.36\% (-0.62\%), and 66.02\% (-2.74\%) on the CIFAR-10, CIFAR-100, and ImageNet datasets, respectively, using the ResNet-18 model with only minimal accuracy loss. 
\end{itemize}

\section{Preliminaries}

\subsection{Memory Access}

In modern computing, a significant challenge is the high energy consumption caused by frequent memory accesses\cite{Energy_Problem}, which is particularly severe in deep learning due to extensive data and computation requirements. Despite the benefits of CiM systems, significant energy is still expended in transferring the intermediate activation data between the CiM macro and cache memory. For example, a 16-bit MAC operation requires only 0.075 pJ of energy, whereas accessing a 512 KB SRAM cache incurs an energy cost of 30.375 pJ\cite{Interstellar}, resulting in a $400\times$ disparity. In an 8-bit ImageNet tasks using the ResNet-50 model, we estimate that SRAM cache read and write operations of the activations alone can consume up to 394 $\mu$J, which is comparable to the 405 $\mu$J MAC energy expenditure. Moreover, loading model parameters into the processor, necessitating DRAM accesses, can cost as much as 200 pJ per access, making weight parameter loading the major energy-draining activity in CiM accelerators\cite{Bit-Parallel,Scalable_Programmable_CiM}. 
One approach to mitigate this issue is through DNN model compression and sparse matrix representation processing\cite{Deep_Compression}, which reduce storage requirements and enable DNN parameters to fit entirely within on-die SRAM, eliminating memory access and enhancing efficiency\cite{EIE}. However, these techniques have received limited attention in CiM research, highlighting an urgent need for developing a specialized computing approach to tackle the memory access issues in CiMs.

\subsection{Approximate Computation of DNN}

DNN models exhibit a remarkable resilience to computational inaccuracies, providing opportunities for resource optimization through approximate computation\cite{Approximate_Survey}. This capability enables the energy reduction of digital accelerators through the application of approximate multiplication, thereby improving inference efficiency\cite{Approximate_Mult_Analysis}. Selecting the most suitable approximate multiplier can lead to substantial energy savings with only minimal impact to model performance\cite{ALWANN}. With transformer models becoming increasingly prevalent, novel approaches have been developed to boost efficiency in processing attention layers, including the use of a big-exact-small-approximate strategy to efficiently manage weakly-related tokens and reduce unnecessary computation of near-zero values\cite{BESA_Transformer}. In the domain of CiM systems, the adoption of approximate computation methods is also gaining attention. Incorporating an approximate adder tree into the Digital-CiM (D-CiM) array improves hardware efficiency while maintaining an RMSE of only 4.03\%\cite{Approximate_Adder}. Although this level of RMSE allows for CIFAR-10 applications, it is still too serious for tasks requiring higher noise tolerance, such as ImageNet, indicating the need for novel methods that achieve both high approximation accuracy and energy efficiency.

\subsection{Hybrid Computation}

CiM architectures can be classified into two types: D-CiM and Analog-CiM (A-CiM)\cite{A-CiM_D-CiM_Benchmark}. D-CiM systems utilize digital logic for dot product (DP) operations within memory arrays and aggregate these DPs using a digital adder tree\cite{D-CiM_Macro,Colonnade}. While digital logic ensures error-free computation, it compromises energy and area efficiency. On the contrary, A-CiM architectures achieve DP accumulation in the charge or current domain, with an ADC for voltage level translation\cite{Segment_Bitline_CiM,Cap_Reconfig_CiM,IMPACT}. A-CiM systems offer high energy efficiency through low-power analog computations, but are more susceptible to errors from device variability and noise. Recent developments in CiM systems incorporate a hybrid architecture that combines the advantages of both D-CiM and A-CiM through hybrid computation\cite{Hybrid_Analog_Digital_CiM}. This approach, grounded in approximate computing principles, assigns higher-order MSB cycles to the digital domain and lower-order LSB cycles to the analog domain, thus enables both high inference quality and low energy consumption. 

Some processors integrate both digital and analog cores within a single system, allocating different DNN layers to either core for processing\cite{DIANA}. Other designs incorporate D-CiM and A-CiM within a H-CiM macro to maximize computation efficiency\cite{FP_H-CiM,OSA-HCIM}. To mitigate the issues caused by ADC and LSB memory access, we explore the feasibility of constructing a specialized approximate processing core based on statistical inference, leading to a novel architecture that surpasses existing H-CiM designs in both computation and data communication efficiencies.

\subsection{Exploiting Sparsity}

Exploiting sparsity, characterized by the zero-valued elements within DNN weights and activations, presents an opportunity for enhancing computational speed and reducing energy consumption. Digital accelerators leverage this characteristic by detecting zero vales and applying gating logic to bypass unnecessary computations and data accesses\cite{Eyeriss}. In CiM systems, the focus has shifted towards bit-level sparsity, where detecting sparse input activations can significantly eliminate redundant circuit operations or ADC conversion cycles\cite{Bit-Parallel,Bit-Flexible_Macro,Z-PIM,Sparsity_RRAM}. Efforts to enhance bit-level sparsity in DNN computations further reduce insignificant operations, improving the operational efficiency of CiM systems\cite{ISSA,Bit-Transformer}. While sparsity has traditionally been regarded as merely an auxiliary tool for improving computational efficiency, our PAC method takes a novel approach by applying bit-level sparsity directly to approximate MAC computations, realizing an energy-efficient but precise computing approach. Moreover, developing a system fully centered around sparsity not only enhances computational efficiency but also effectively addresses the challenges related to memory access, which have been difficult to tackle within other low-power computing techniques.

\begin{figure}[htbp]
\centering
\includegraphics[width=0.47\textwidth]{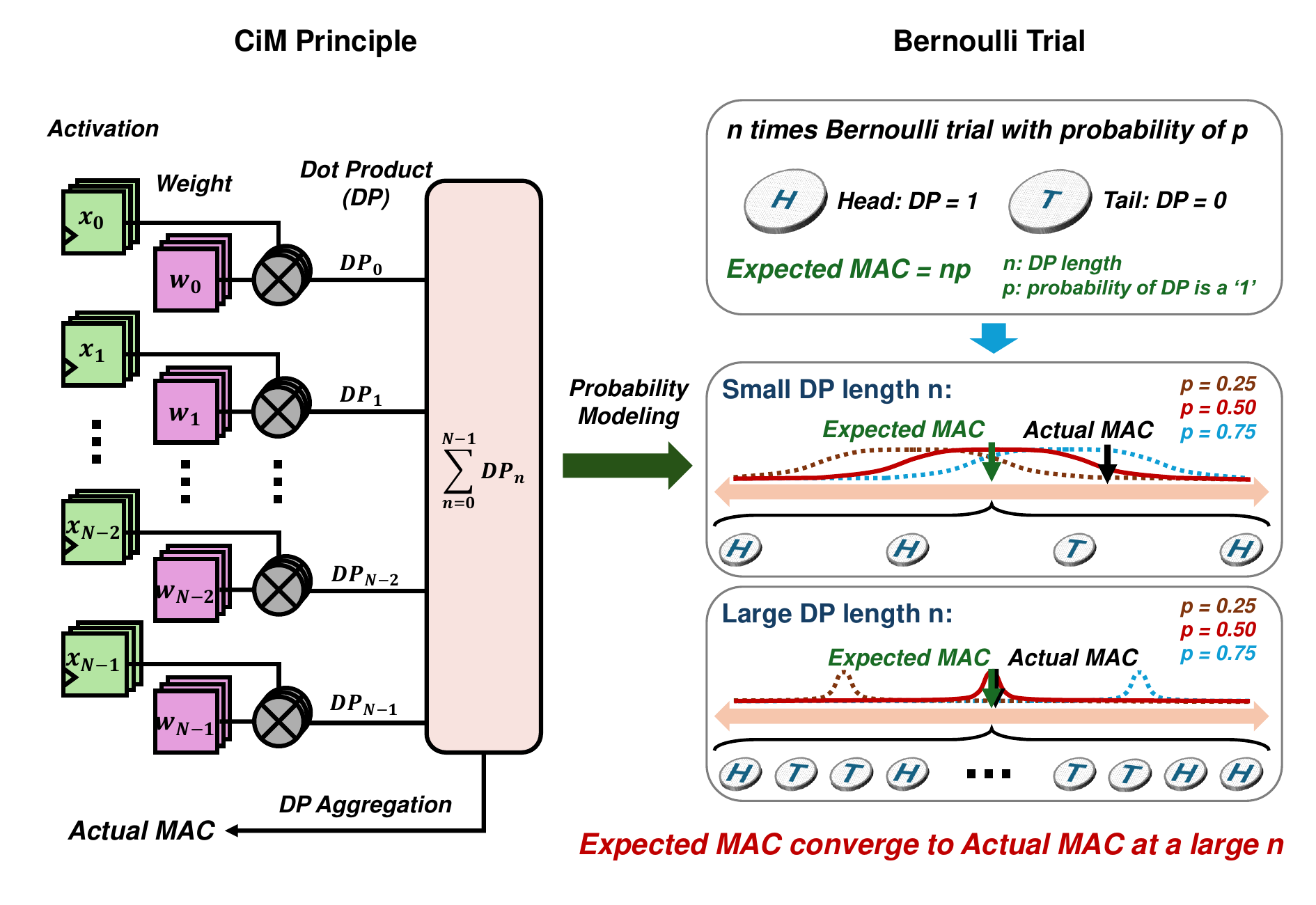}
\caption{Probabilistic Approximate Computation: Model CiM operations using principles of statistical inference.}
\label{Fig_PAC}
\end{figure}

\section{Probabilistic Approximate Computation}

\begin{figure*}[htbp]
  \centering
  \begin{subfigure}[b]{0.32\textwidth}
    \includegraphics[width=\textwidth]{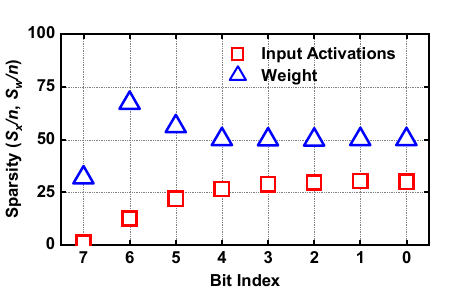}
    \caption{}
    \label{Fig_Sparsity_Range}
  \end{subfigure}
  % \hfill
  \begin{subfigure}[b]{0.32\textwidth}
    \includegraphics[width=\textwidth]{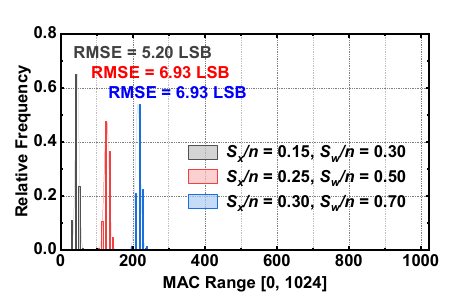}
    \caption{}
    \label{Fig_Distribution}
  \end{subfigure}
  % \hfill
  \begin{subfigure}[b]{0.32\textwidth}
    \includegraphics[width=\textwidth]{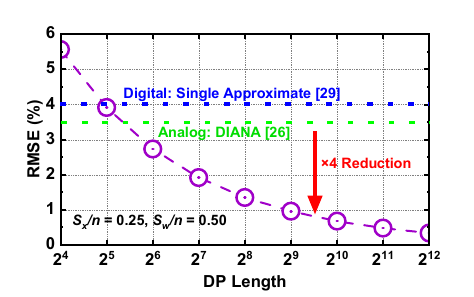}
    \caption{}
    \label{Fig_Error}
  \end{subfigure}
  \caption{Approximate Error Analysis: (a) Weight/activation sparsity dependence across each bit index. (b) Distribution of actual MAC operations for typical weight/activation sparsity combinations. (c) RMSE variation across different DP lengths.}
  \label{Fig_Error_Analysis}
\end{figure*}

\subsection{Concept Formulation}

\noindent \textbf{Computing Concept.} We first review the fundamental principles of CiM systems and then introduce our novel PAC method. CiM accelerators primarily perform MAC operations and compute output activations by the following equation:

\begin{equation}
O = \mathbf{x} \cdot \mathbf{w} = \sum_{p = 0}^{P - 1} \sum_{q = 0}^{Q - 1} {2}^{p + q} \sum_{n = 0}^{N - 1} x_{n}[p] w_{n}[q]
\label{Eq_MAC}
\end{equation}

\noindent Here, $\mathbf{x}$ and $\mathbf{w}$ represent the N-dimensional input activation vector and weight vector, respectively, both in \textsc{UINT} format with bit widths P and Q. In both D-CiM and A-CiM systems, the primary computational activity is centered around generating DPs within local memory cells, followed by summing instances where DP equals `1' in a CiM column, as detailed in Eq. \ref{Eq_MAC}. Moving beyond traditional techniques, we propose a new approach that leverages statistical inference and exploits bit-level sparsity for approximation. Assuming the occurrence of DP equaling `1' follows a Bernoulli distribution with probability $p$ over length $n$, we consider the local DP generation a Bernoulli trial, depicted in Fig. \ref{Fig_PAC}, with the MAC output following a binomial distribution $MAC \sim B(n, p)$. By modeling the DP generation to a probabilistic expression and knowing the probability of DP being `1', we can approximate the MAC output via point estimation, calculating the expected value as $E(MAC) = np$. As $n$ increases, the expected MAC output progressively aligns with the actual value, affirming the accuracy of this approximation method in MAC computations. In common \textsc{AND} logic CiM scenarios, we define $p$ as follows:

\begin{equation}
P(DP = 1) = P(w = 1 \cap x = 1) = P(w = 1) \cdot P(x = 1)
\label{Eq_DP_Probability}
\end{equation}

\noindent In this context, $P(DP = 1)$, which stands for $P(wx = 1)$, indicates the probability that the local DP equals `1'. This value is determined by multiplying the probabilities of the weight and activation bits being `1', under the assumption that these bits are independent and identically distributed. Utilizing the total count of `1's for each bit, we express these probabilities via bit-level sparsity $S_{w}/n$ and $S_{x}/n$, where $S_{w}$ and $S_{x}$ indicate the total `1's for weights and activations throughout the DP sequence. Thus, the expected MAC output for \textsc{AND} logic CiM systems simplifies to:

\begin{equation}
E(MAC) = S_{x} \cdot S_{w} / n
\label{Eq_PAC}
\end{equation}

\noindent \textbf{Data Encoding.} To derive $S_{w}$ and $S_{x}$ for PAC, bit-level sparsity encoding is applied to weights and activations. Taking a CNN as an example, for a multi-bit original activation $\mathbf{x}$ and weight $\mathbf{w}$ with a $bit \times channel$ format, the sparsity for each bit index is determined by the total number of `1's across the channel. Illustrated in Fig. \ref{Fig_Concept}, this encoding compresses the channel dimension per bit index, converting the data representation to $bit \times 1$. For instance, an 8-bit tensor for weight $\mathbf{w}$ or activation $\mathbf{x}$, sized at $8 \times 128$ bits, undergoes encoding to become $8 \times 7$ bits. This method significantly compresses the data: from 1024 bits originally to just 56 bits, achieving a 95\% compression ratio and markedly enhancing data transfer efficiency.

\begin{table}[htbp]
\centering
\begin{threeparttable}
\caption{Error of State-of-the-Art Approximate Methods}
\label{Table_RMSE}
\begin{tabular}{ccccc}
\Xhline{1.5pt}
& \textbf{\thead{\small ISSCC \\ 2022\cite{Approximate_Adder}}} & \textbf{\thead{\small ISSCC \\ 2022\cite{DIANA}}} & \textbf{\thead{\small ASP-DAC \\ 2024\cite{OSA-HCIM}}} & \textbf{\thead{\small This Work}} \\
\Xhline{1.5pt}
\textbf{\thead{\small Method}} & \thead{\small Approximate} & \thead{\small Analog} & \thead{\small Analog} & \textbf{\thead{\small Sparsity}} \\
\hline
\textbf{\thead{\small RMSE (\%)}} & \thead{\small 4.0/6.8\tnote{a}} & \thead{\small 3.5-4.8\tnote{b}} & \thead{\small 8.5\tnote{c}} & \textbf{\thead{\small 0.3-1.0\tnote{d}}} \\
\Xhline{1.5pt}
\end{tabular}
\begin{tablenotes}
\small
\item[a] RMSE (\%) on single/double-approximate hardware.
\item[b] Error (\%) minimum at best operating point reported in Ref. \cite{DIANA_JSSC}.
\item[c] RMSE (\%) calculated by Macro spec with quantization error.
\item[d] RMSE (\%) with DP length from 512 to 4096.
\end{tablenotes}
\end{threeparttable}
\end{table}

\noindent \textbf{Efficient H-CiM System.} According to Eq. \ref{Eq_PAC}, our PAC methodology transforms computation into the sparsity domain, effectively converting the traditional n-fold product-sum computation into a single multiply-divide operation, significantly conserving the computational energy. Moreover, unlike traditional methods that read each bit from a vector individually for computation, PAC requires just one read to obtain the sparsity information, substantially reducing memory access power. Given that the MSB part computation is crucial and nearly noise-intolerant, we incorporate PAC within deterministic digital computation and aim to reformulate Eq. \ref{Eq_MAC}, crafting an H-CiM architecture:

\begin{equation}
\begin{aligned}
O = \mathbf{x} \cdot \mathbf{w} &\approx \sum_{(p, q) \in \mathbb{D}} {2}^{p + q} \sum_{n = 0}^{N - 1} x_{n}[p] w_{n}[q] \\
&+ \sum_{(p, q) \in \mathbb{A}} {2}^{p + q} (S_{x}[p] \cdot S_{w}[q] / N)
\end{aligned}
\label{Eq_Hybrid_Computation}
\end{equation}

\noindent In this equation, $(p, q)$ denotes a binary MAC cycle correlating bit indices of activation and weight, with $\mathbb{D}$ and $\mathbb{A}$ representing deterministic and approximate binary MAC cycle subsets, respectively. The terms $S_{x}[p]$ and $S_{w}[q]$, representing the summation of instances where the p-th activation and q-th weight bits are `1', transforms complex vector calculations into streamlined scalar operations, thus introducing an unique approach to conserve computational energy.

\subsection{Approximate Error Analysis}

Next, we evaluate the approximate errors associated with the PAC methodology. We quantized both the ResNet-18 model and CIFAR-100 input activations into \textsc{UINT8} format to observe the sparsity at each bit index. As depicted in Fig. \ref{Fig_Error_Analysis}(a), weight sparsity fluctuates between 0.25 and 0.7, whereas input activation sparsity varies from 0 to 0.3, highlighting the dynamic nature of sparsity across bit indices. To quantify the RMSE of our PAC approach, we simulated a CiM column with a DP length of 1024. By randomly generating binary weight and activation bits with three specific sparsity levels, we recorded the actual MAC outputs and compared them against the expected values derived via the PAC methodology. This experiment was conducted over 100K iterations, demonstrated that the actual MAC outputs closely align with the expected values as shown in Fig. \ref{Fig_Error_Analysis}(b), with a RMSE of around 6 LSB. This result implies that the PAC method ensures a deviation of less than 0.6\% in over 68\% of computations for a DP length of 1024. Furthermore, as illustrated in Table \ref{Table_RMSE}, the PAC methodology surpasses the accuracy of existing approximate computing techniques by fourfold, utilizing merely a simple multiply-divide operation while also achieving greater power efficiency. According to the law of large numbers, we expect the accuracy of the PAC method to improve with the extension of DP lengths. Fig. \ref{Fig_Error_Analysis}(c) depicts an analysis of the RMSE (\%) trend over DP lengths ranging from 16 to 4096, where the accuracy of the PAC method surpasses that of alternative approximation methods at a DP length of 64, with the RMSE (\%) consistently decreasing in proportion to $n^{-1/2}$ at greater DP lengths\cite{SC_Survey,RRAM_SC_CiM}. Given that DP lengths in typical \textsc{CONV} layers range from $3 \times 3 \times 64$ to $3 \times 3 \times 512$, and \textsc{LINEAR} layers span from 512 to 4096, our PAC can approximate MAC outputs with an RMSE below 1\%. This positions our PAC technique as a superior alternative to existing approximation methods, paving the way for both accurate and efficient CiM system.

\begin{figure}[htbp]
\centering
\includegraphics[width=0.5\textwidth]{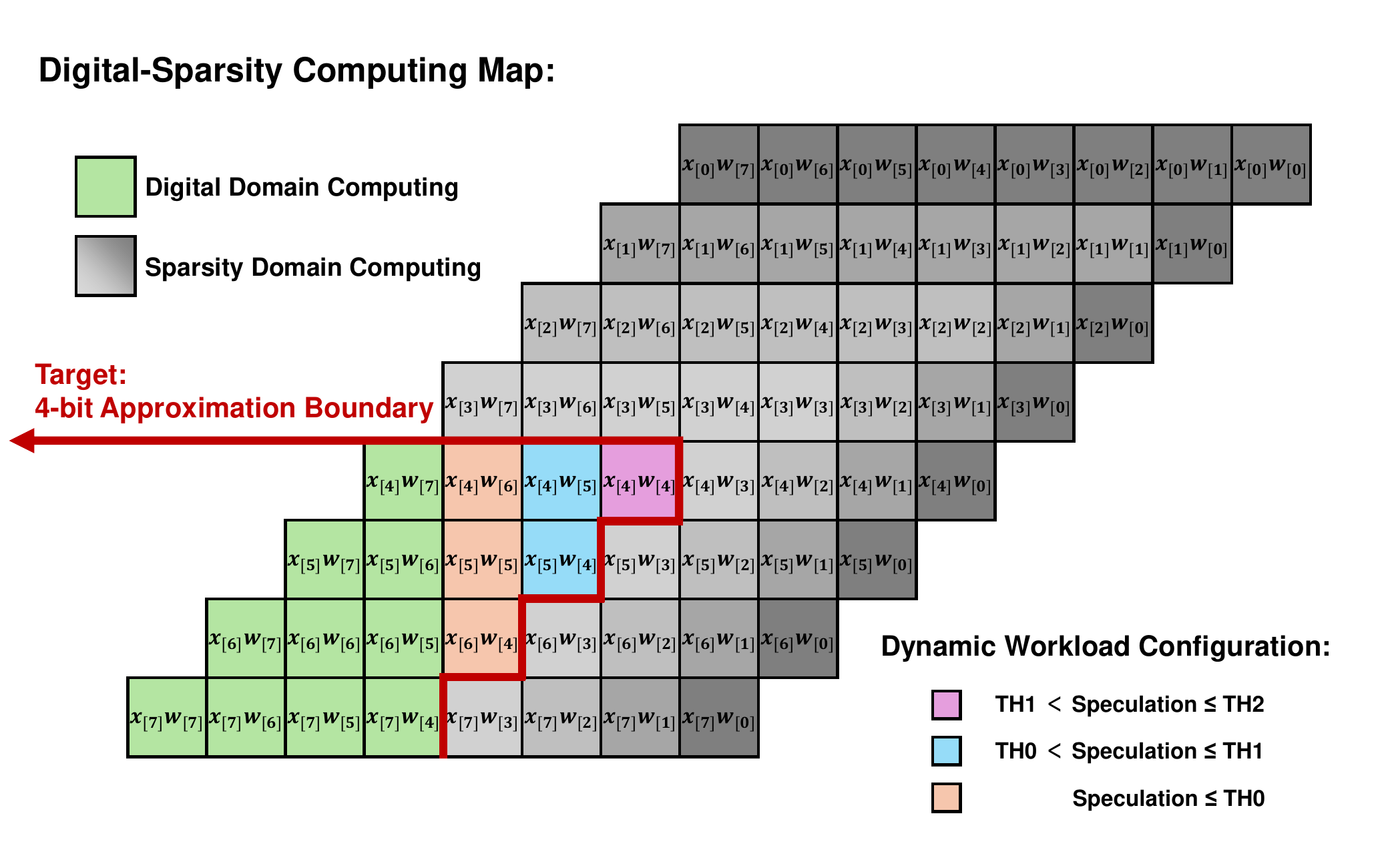}
\caption{Computing map of the PACiM architecture.}
\label{Fig_Computing_Map}
\end{figure}

\vspace*{-3mm}

\section{Architecture}

\subsection{Computing Map}

This section introduces the computing map that configures bit-serial computing cycles to digital and sparsity domains in our PACiM architecture, as described in Eq. \ref{Eq_Hybrid_Computation}. Unlike traditional H-CiM systems that divide computing cycles based on bit-shift orders , our PACiM architecture employs a distinct operand-based approximation, as shown in Fig. \ref{Fig_Computing_Map}. This approach allows for the removal of memory columns that store LSB weights in CiM macros by leveraging bit-level sparsity in the sparsity domain computing. Operand-based approximation enables simultaneous maximization of computational efficiency and minimization of area overhead. Our PACiM architecture specifically targets a 4-bit approximation for 8-bit DNN tasks, and this eliminates four memory columns dedicated to the LSB weights. Consequently, the D-CiM computing cycles are significantly reduced from 64 to 16 as shown in Fig. \ref{Fig_Computing_Map}, where the gray squares represent the approximate computing cycles transferred to the sparsity domain.

\subsection{Architecture Overview}

\begin{figure*}[htbp]
\centering
\includegraphics[width=\textwidth]{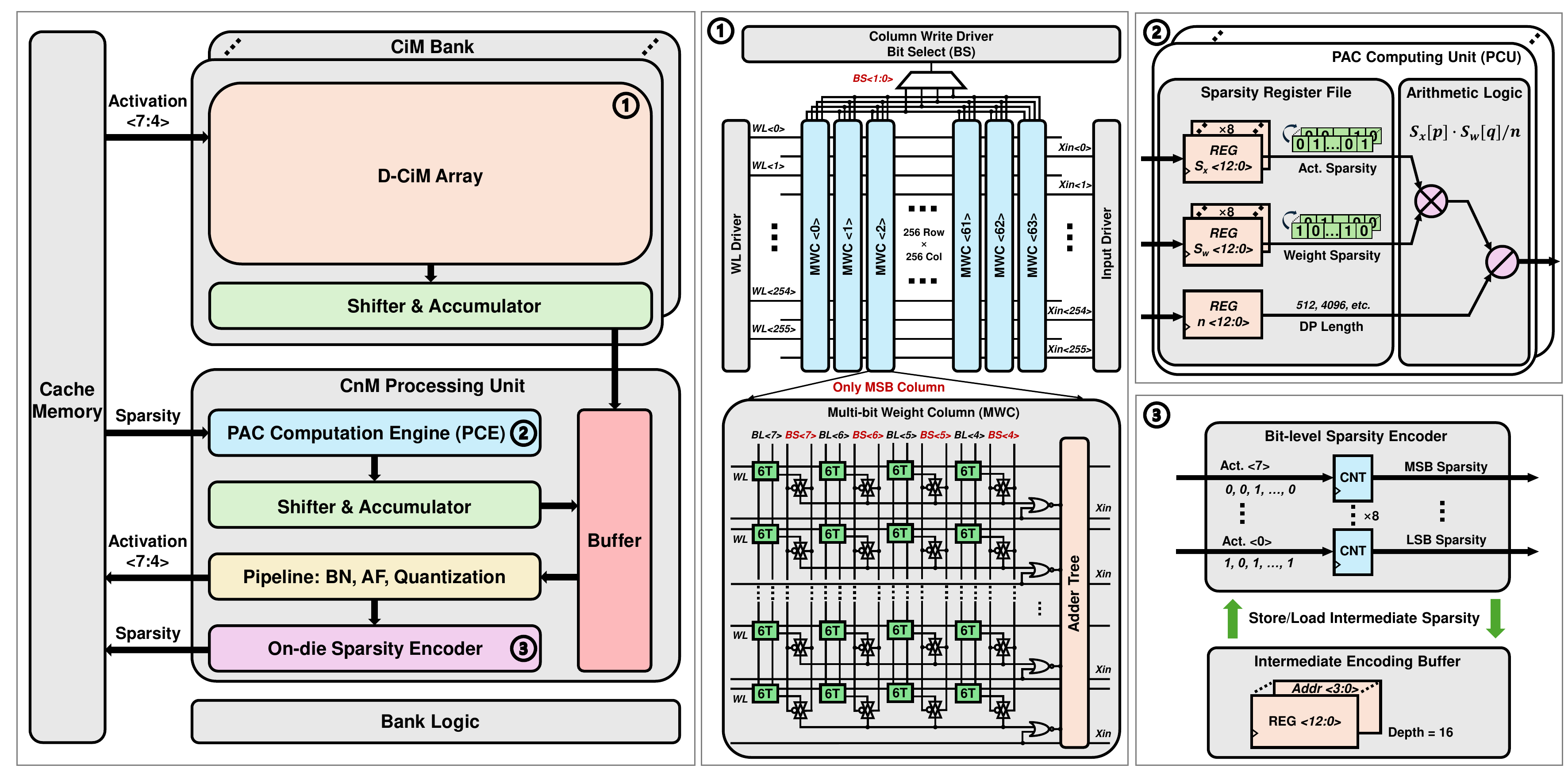}
\caption{PACiM Architecture: D-CiM banks for deterministic computing and CnM processing unit for PAC processing.}
\label{Fig_Architecture}
\end{figure*}

We present the PACiM architecture, illustrated in Fig. \ref{Fig_Architecture}, which integrates multiple critical components. This architecture includes multiple D-CiM banks for deterministic computing in the digital domain, a CnM processing unit for handling approximate computations in the sparsity domain, and dedicated bank logic for managing system operations. In our PACiM architecture, which targets a 4-bit approximation in 8-bit DNN tasks, only the 4-bit MSB computation of weights and activations are processed using deterministic D-CiM. The remaining LSB computations are loaded as sparsity data and handled entirely in the sparsity domain. Specifically, weights are pre-processed offline and converted into a 4-bit MSB format, integrated with bit-level sparsity. On the other hand, activation sparsity is processed online during DNN inference, while the 4-bit LSB activation is completely discarded. This approach effectively eliminates the need for weight updates and activation loading within the CiM array, reducing DRAM access and cache read/write activities by nearly 50\%.

The computing process initiates with 4-bit MSB activations being sent from the cache to the D-CiM banks for bit-serial computation. These activations are processed across multiple columns, with binary MAC outputs undergoing bit-shifting and accumulation before being transferred in parallel to the buffer. Simultaneously, the CnM processing unit manages sparsity computations, with the PAC computation engine (PCE) for sparsity domain computation, accumulating approximate binary MAC results and sequentially transferring them to the buffer. Finally, the buffer integrates results from both the D-CiM banks and the PCE, channeling them through a pipeline of batch normalization (BN), activation function (AF), and quantization. The architecture recycles the 4-bit MSB portion of activations back to the cache for subsequent layer computations, while the on-die sparsity encoder compresses each bit index into a compact, low-dimensional sparsity representation before being sent back to the cache, significantly enhancing data communication efficiency.

\subsection{D-CiM Array}
Fig. \ref{Fig_Architecture} \textcircled{1} details the structure of $256 \times 256$ cells D-CiM array similar to Ref. \cite{D-CiM_Macro}, featuring 64 multi-bit weight columns (MWCs), wordline (WL)/input drivers, column write drivers, and bit select (BS). Each MWC is designed to store 4-bit MSB weights that are indexed to corresponding memory columns. A key feature is the integration of a transmission gate, controlled by the BS, on the $\overline{Q}$ side of each 6T SRAM cell. This design allows for the selective introduction of weight bits into one side of the \textsc{NOR} gate array during operation, while input activations are simultaneously loaded on the other side, ultimately aggregating the DPs in the adder tree for efficient bit-serial MAC computations. Importantly, the adoption of the PAC method allows for the elimination of memory columns for LSB storage.

\subsection{PAC Computation Engine}

Fig. \ref{Fig_Architecture} \textcircled{2} shows the circuit design of the PCE, consisting of multiple PAC computing units (PCUs). Each PCU includes a sparsity register file to store sparsity data and arithmetic logic for conducting PAC computations in accordance with Eq. \ref{Eq_PAC}. These units handle sparsity domain computations, aggregating the results in an accumulator for each MAC cycle. Although the PCE operates independently from the CiM banks, it is crucial that the number of PCUs matches the throughput of the CiM banks to ensure optimal system utilization. Regarding circuit operations, weight sparsity data are loaded into corresponding registers, and input sparsity registers is refreshed dynamically from the cache. This configuration enables the PCUs to function in a weight-stationary fashion during sparsity domain computations, thus optimizing operational power. In our implementation with 65 nm CMOS, each PCU with its corresponding accumulator, including all register files and arithmetic logic, occupies only 8640 ${\mu m}^2$ of area while achieving a 12-fold improvement in energy efficiency compared to D-CiM, largely due to the simplification of MAC computations by the PAC method.

\begin{figure}[htbp]
    \centering
    \begin{subfigure}[b]{0.49\columnwidth}
        \includegraphics[width=\textwidth]{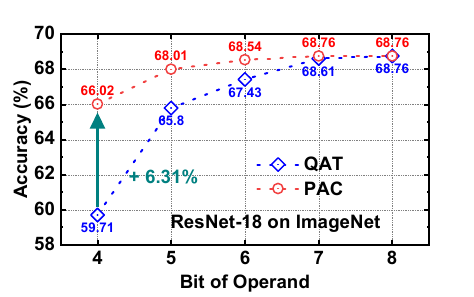}
        \caption{}
        \label{Fig_Bit_Dependence}
    \end{subfigure}
    \hfill
    \begin{subfigure}[b]{0.49\columnwidth}
        \includegraphics[width=\textwidth]{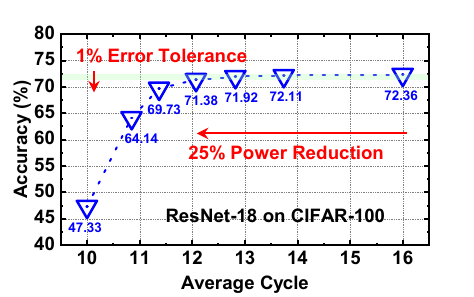}
        \caption{}
        \label{Fig_Dynamic_Configuration}
    \end{subfigure}
    \caption{Evaluation of DNN Inference Accuracy: (a) PAC versus QAT. (b) Dynamic workload configuration.}
    \label{Fig_Accuracy}
\end{figure}

\vspace*{-3mm}

\subsection{On-die Sparsity Encoder}

Fig. \ref{Fig_Architecture} \textcircled{3} unveils the composition of on-die sparsity encoder, designed to convert 8-bit activations from pipeline output into sparsity format. It utilizes eight counters to track the occurrences of `1's within each bit of the activations. The encoding strategy is adapted for both \textsc{CONV} and \textsc{LINEAR} layers. 

\noindent \textbf{\textsc{CONV} Layers.} For \textsc{CONV} layers, where convolution kernels are arranged along MWCs, activation broadcasting across the array produces output activations within identical pixels across different channels. Therefore, pixel-wise encoding across all channels offers the most optimal approach for encoding output activations.

\noindent \textbf{\textsc{LINEAR} Layers.} In Linear layers, given that activations from input layer are recomputed for every output neuron, layer-wise encoding emerges as the most efficient strategy, encapsulating each neuron within the layer.

\noindent \textbf{Intermediate Encoding Buffer.} In a single-bank system design, when the extensive MAC operations of an output activation cannot be entirely accommodated within the MWCs, encoding on a specific pixel or layer may be disrupted during weight updates in the D-CiM array. Consequently, an intermediate encoding buffer is employed to temporarily store the current sparsity. The state of the sparsity counter is configurable, allowing adaptability by loading sparsity data from the buffer. The intermediate encoding buffer functions as a temporary storage for the current sparsity when the MAC operations of an output activation exceed the capacity of the MWCs. It enables the sparsity counter to resume encoding from its previous state by loading the stored sparsity data from the buffer.

\begin{figure*}[htbp]
  \centering
  \begin{subfigure}[b]{0.32\textwidth}
    \includegraphics[width=\textwidth]{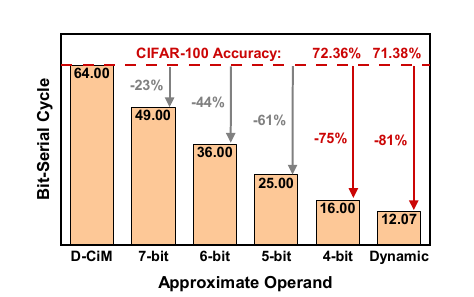}
    \caption{}
    \label{Fig_Computation_Cycle}
  \end{subfigure}
  % \hfill
  \begin{subfigure}[b]{0.32\textwidth}
    \includegraphics[width=\textwidth]{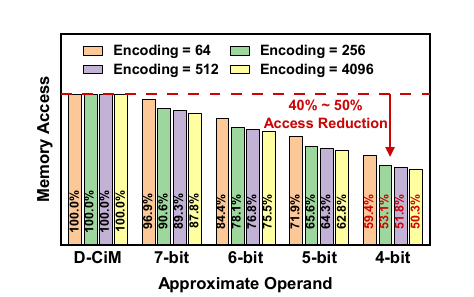}
    \caption{}
    \label{Fig_Memory_Access}
  \end{subfigure}
  % \hfill
  \begin{subfigure}[b]{0.32\textwidth}
    \includegraphics[width=\textwidth]{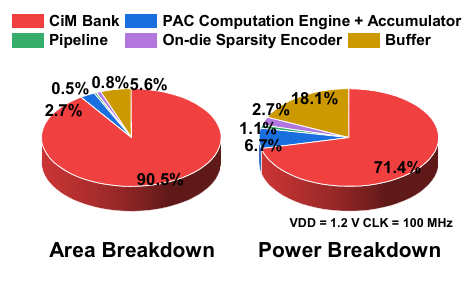}
    \caption{}
    \label{Fig_Breakdown}
  \end{subfigure}
  \caption{System Analysis of PACiM: (a) Bit-serial cycle and (b) Memory access. (c) Area and power breakdown in our design.}
  \label{Fig_System_Analysis}
\end{figure*}

\noindent \textbf{Tiling Multiple Banks.} By tiling multiple banks through a carefully designed network of connections and control logic to form a larger system, the need for these intermediate encoding buffers can be eliminated. In a multi-bank system, the weight updates can be scheduled in a way that avoids disrupting the encoding process within a single bank. This allows for continuous encoding of activations without the need for temporary storage in the intermediate encoding buffers.

\section{Dynamic Workload Configuration}

Unlike traditional CiM architectures, our PACiM architecture uniquely identifies the sparsity of activations before broadcasting inputs across the CiM array. This distinctive feature motivates us to maximize the utilization of activation sparsity, thereby further improving energy efficiency. According to Ref. \cite{OSA-HCIM}, a saliency-aware dynamic configuration can be implemented based on the significance of each image pixel. In our PACiM architecture, we can estimate the saliency of each output activation by examining the sparsity of corresponding input activations. Intuitively, a denser bit distribution in the input activations indicates that the computation is likely more salient, suggesting a higher output activation. As a result, we can speculate on the MAC output using the following equation:

\begin{equation}
SPEC = \sum_{p = 0}^{P - 1} {2}^{p} \cdot S_{x}[p]
\label{Eq_Speculation}
\end{equation}

\noindent Eq. \ref{Eq_Speculation} provides a weighted sum of bit-level activation sparsity, which serves as an indicator of MAC output magnitude. By speculating the magnitude of MAC outputs, we selectively reduce bit-serial cycles for MACs that minimally impact DNN inference outcomes. For speculations that fall below a certain value, we modulate the boundary between digital and sparsity computing, shifting operations more toward the sparsity domain. This approach is particularly effective in tasks such as CIFAR-10 and CIFAR-100, where DNN models are more tolerant of computational noise. This speculation mechanism is integrated into the bank logic, facilitating a dynamic workload configuration that modulates system operation. Specifically, we normalize Eq. \ref{Eq_Speculation} and apply a threshold set [\textsc{TH0}, \textsc{TH1}, \textsc{TH2}] to dynamically adjust the digital-sparsity computing boundary. Speculations exceeding \textsc{TH2} are deemed crucial for inference, necessitating the full computation of 16 bit-serial cycles in a 4-bit approximation context. Moderate speculations trigger an incremental transfer of cycles to the sparsity domain, as shown in Fig. \ref{Fig_Computing_Map}. Upon speculations not surpassing \textsc{TH0}, the impact on inference is consistently marginal, which facilitates a reduction in bit-serial cycles to the optimal minimum of 10.

\begin{table}[htbp]
\centering
\caption{Inference Accuracy | Loss on 4-bit Approximation.}
\label{Table_Accuracy}
\begin{tabular}{@{}lcccc@{}}
\Xhline{1.5pt}
 & \textbf{CIFAR-10} & \textbf{CIFAR-100} & \textbf{ImageNet} \\
\Xhline{1.5pt}
\textbf{ResNet-18} & 93.85\% | -0.62\% & 72.36\% | -0.62\% & 66.02\% | -2.74\% \\
\hline
\textbf{ResNet-50} & 93.21\% | -1.02\% & 72.65\% | -1.04\% & 75.98\% | -3.38\% \\
\hline
\textbf{VGG16-BN}  & 94.29\% | -0.66\% & 75.39\% | -0.69\% & 71.59\% | -1.31\%  \\
\Xhline{1.5pt}
\end{tabular}
\end{table}

\begin{table}[htbp]
\centering
\begin{threeparttable}
\caption{1b/1b Energy Efficiency with Supply at 0.6/1.2 V}
\label{Table_TPOSW}
\begin{tabular}{cccc}
\Xhline{1.5pt}
 & \textbf{D-CiM} & \textbf{PCU + Acc. } & \textbf{PACiM} \\
\Xhline{1.5pt}
\textbf{TOPS/W} & 235.01/58.72 & 2945.92/736.48 & 1170.28/292.57\ \\
\Xhline{1.5pt}
\end{tabular}
\end{threeparttable}
\end{table}

\vspace*{-3mm}

\section{Simulation Results}

\begin{table*}[htbp]
\centering
\caption{Performance Comparison with State-of-the-Art CiM Design.}
\label{Table_Comparison}
\begin{threeparttable}
\begin{tabular}{ccccccc}
\Xhline{1.5pt}
 & \textbf{\thead{\small ISSCC 2021\cite{D-CiM_Macro}}} & \textbf{\thead{\small ISSCC 2022\cite{Approximate_Adder}}} & \textbf{\thead{\small ISSCC 2022\cite{DIANA}}} & \textbf{\thead{\small ASP-DAC 2024\cite{OSA-HCIM}}} & \textbf{\thead{\small ISSCC 2024\cite{Cap_Reconfig_CiM}}} & \textbf{\thead{\small This Work}} \\
\Xhline{1.5pt}
\textbf{\thead{Type}} & \thead{Digital} & \thead{Approximate} & \thead{Digital-Analog} & \thead{Digital-Analog} & \thead{Analog} & \textbf{\thead{Digital-Sparsity}} \\
\hline
\textbf{\thead{Node}} & \thead{22 nm} & \thead{28 nm} & \thead{22 nm} & \thead{65 nm} & \thead{65 nm} & \textbf{\thead{65 nm}} \\
\hline
\textbf{\thead{Precision}} & \thead{4b/4b, 8b/8b} & \thead{4b/1b} & \thead{A: 2b/2b, D: 8b/8b} & \thead{8b/8b} & \thead{4-8b} & \textbf{\thead{8b/8b}} \\
\hline
\textbf{\thead{Supply (V)}} & \thead{0.72} & \thead{0.5} & \thead{0.55} & \thead{0.6} & \thead{0.6} & \textbf{\thead{0.6}} \\
\hline
\textbf{\thead{Peak TOPS/W\tnote{*}}} & \thead{163.13} & \thead{2219/992} & \thead{74.88} & \thead{245.12–370.56} & \thead{4094/818} & \textbf{\thead{1170.28}} \\
\hline
\textbf{\thead{CIFAR-10 Acc.\tnote{*}}} & N/A & \thead{86.96\%/90.41\%} & \thead{89\%} & \thead{N/A} & \thead{91.7\%/95.8\%} & \textbf{\thead{93.85\%}} \\
\hline
\textbf{\thead{CIFAR-100 Acc.\tnote{*}}} & \thead{N/A} & \thead{N/A} & \thead{N/A} & \thead{67.4\%–72.1\%} & \thead{N/A} & \textbf{\thead{72.36\%}} \\
\hline
\textbf{\thead{ImageNet Acc.\tnote{*}}} & \thead{N/A} & \thead{N/A} & \thead{64\%} & \thead{65.2\%–70.8\%} & \thead{N/A} & \textbf{\thead{66.02\%}} \\
\hline
\textbf{\thead{Mem. Access Red.}} & \thead{NO} & \thead{NO} & \thead{NO} & \thead{NO} & \thead{NO} & \textbf{\thead{40\%-50\%}} \\
\Xhline{1.5pt}
\end{tabular}
\begin{tablenotes}
\small
\item[*] TOPS/W normalized to 1b/1b MAC in 65 nm CMOS by the bit-serial cycles and node feature capacitance. Accuracy evaluated with ResNet-18/20 models.
\end{tablenotes}
\end{threeparttable}
\end{table*}

\subsection{Software Simulation on DNN Accuracy}

To assess the inference accuracy of our PACiM architecture, we developed a PyTorch-based simulation framework that accurately reflects bitwise operations of CiM. This framework decomposes weights and input activations into binary tensors, closely aligning bit-serial MAC computations with the behavior expected in actual circuits. Given that our PACiM system is fully digital and all errors originate from the statistical noise in bit distribution, this approach allows us to evaluate the inference performance accurately. In our system configuration, the initial $3 \times 3 \times 3$ \textsc{CONV} layer uses standard D-CiM for accurate feature extraction\cite{DIANA}, whereas all subsequent \textsc{CONV} and \textsc{LINEAR} layers apply our 4-bit approximate PAC method. The DNN models are first pre-trained through 8-bit \textsc{UINT} quantization-aware training (QAT), followed by fine-tuning under progressively augmented Gaussian noise. We found that directly imposing a high level of Gaussian noise challenges the convergence process and diminishes model accuracy. However, beginning with a good intialization enables the models to demonstrate superior noise tolerance relative to conventional noise-aware training techniques.

We investigated the inference accuracy of our PACiM architecture using various approximate operands on the ImageNet benchmark with the ResNet-18 model, as illustrated in Fig. \ref{Fig_Accuracy}(a). The performance was also compared to that of models trained via QAT at corresponding bit widths, which were directly adjusted to accommodate a lower bit precision. The PAC-based approximation of an 8-bit DNN model surpasses the performance of conventional QAT, particularly in the 4-bit QAT, where insufficient expressiveness leads to an accuracy drop to 59.71\%. The PAC method compensates for this by approximating the behavior of the 8-bit model, raising accuracy to 66.02\%. In contrast, the memory cell columns that store LSB bits can be completely discarded in the 4-bit approximation of PAC, significantly reducing the bit cell area by half. Table \ref{Table_Accuracy} summarizes the accuracy of 4-bit PAC approximation across different models and tasks. In CIFAR-10 and CIFAR-100, the accuracy loss is negligible, remaining under 1\% compared to the 8-bit QAT baseline. This minor drop is primarily from the procedure of noise fine-tuning, indicating that the PAC method is effective for real-world DNN tasks. However, in ImageNet, which demands higher computational precision, the accuracy loss is slightly greater but still remains nearly 3\%. For applications where this level of accuracy loss is unacceptable, switching to a 5-bit approximation effectively eliminates the loss to <1\%, while still maintaining high energy efficiency. Fig. \ref{Fig_Accuracy}(b) demonstrates the potential for dynamic workload configuration to reduce bit-serial cycles in the CIFAR-100 task. By setting the threshold values referring to Eq. \ref{Eq_Speculation}, the average bit-serial cycle can be lowered to 12, achieving an additional 25\% reduction on top of the minimum level in 4-bit approximation, with only a 1\% degradation in accuracy.

\subsection{System Performance}

To evaluate the system performance of our PACiM architecture, we referenced the 4b/4b D-CiM bank specifications reported in Ref. \cite{D-CiM_Macro} and normalized both the area and power consumption to a 65 nm CMOS for our analysis. Additionally, we designed the CnM processing unit to complement the corresponding D-CiM bank. We developed the RTL for the entire CnM processing unit using Verilog, synthesized it with the Synopsys Design Compiler for the 65 nm CMOS, and completed the layout with the Synopsys IC Compiler. This process allowed us to analyze both the area and power consumption. To match the throughput with a 64-accumulator single bank output, we integrated 6 PCUs into the PCE.

The implementation of a 4-bit approximation using PAC in our study successfully reduces the bit-serial cycles for 8-bit DNN tasks by 75\%, shifting these computations into the sparsity domain, as illustrated in Fig. \ref{Fig_System_Analysis}(a). While the D-CiM bank exhibits a binary MAC efficiency of 235.01 TOPS/W, the PCU in conjunction with the accumulator achieves a substantially increased efficiency of 2945.92 TOPS/W, indicating an improvement by a factor of twelve as indicated in Table \ref{Table_TPOSW}. By integrating the entire CnM processing unit into the D-CiM bank, we developed an 8b/8b hybrid computing system that elevates the efficiency to 14.63 TOPS/W (1170.28 TOPS/W when normalized to 1b/1b), approximately five times higher than that of a fully digital system. Fig. \ref{Fig_System_Analysis}(b) highlights that even at a channel length of 64, the 4-bit approximation of PACiM reduces cache memory access by 40\%, with potential reductions up to 50\% in deeper \textsc{CONV} or \textsc{LINEAR} layers. Fig. \ref{Fig_System_Analysis}(c) presents an area and power breakdown for our single bank PACiM system, revealing that the CnM processing unit occupies only 10\% of the system size and consumes 30\% of the total power. Remarkably, the buffer within the CnM processing unit accounts for over half of its area and 70\% of its power consumption. However, most of these buffer can be removed in a multi-bank system. As shown in Table \ref{Table_Comparison}, our PACiM architecture achieves energy efficiency comparable to A-CiM and demonstrates nearly a fourfold improvement over digital-analog H-CiM approaches. While the macro designs in Refs. \cite{Approximate_Adder} and \cite{Cap_Reconfig_CiM} appear more efficient, PACiM supports complex tasks requiring greater precision by approximating wider computation operands. Additionally, PACiM is the first CIM design that replaces the LSB bit transmission with sparsity, leading to a 50\% reduction in memory access and a substantial boost in system efficiency.

\section{Conclusion}

In this paper, we explore a novel PAC methodology for approximating MAC computations and propose a custom-designed sparsity-centric architecture to maximize the benefits of the PAC method. Our PACiM architecture can reduce traditional bit-serial CiM cycles by 81\%, achieving 8b/8b peak efficiency levels of 14.63 TOPS/W, while also reducing memory access by 50\%. Despite these energy efficiency gains, our PACiM system maintains high inference accuracy with marginal loss, achieving accuracy of 93.85\%, 72.36\%, and 66.02\% on CIFAR-10, CIFAR-100, and ImageNet tasks, respectively. 

%%
%% The acknowledgments section is defined using the "acks" environment
%% (and NOT an unnumbered section). This ensures the proper
%% identification of the section in the article metadata, and the
%% consistent spelling of the heading.
\begin{acks}
This research was supported in part by the JST CREST JPMJCR21D2, JSPS Kakenhi 23H00467, Futaba Foundation, Asahi Glass Foundation, and the Telecommunications Advancement Foundation.
\end{acks}

\newpage

%%
%% The next two lines define the bibliography style to be used, and
%% the bibliography file.
\bibliographystyle{ACM-Reference-Format}
\bibliography{reference}
% \printbibliography

%%
%% If your work has an appendix, this is the place to put it.

\end{document}